\pdfoutput=1

\documentclass[pra,10pt,superscriptaddress, footnoteinbib]{revtex4}
\usepackage{amsmath}
\usepackage{latexsym}
\usepackage{amssymb}
\usepackage{graphics,epstopdf}
\usepackage[colorlinks=true, citecolor=blue, urlcolor=blue ]{hyperref}
\usepackage{epsf,graphics,graphicx}

\newcommand{\n}{\noindent}

\newcommand{\ed}{\end{document}}
\newcommand{\beq}{\begin{equation}}
\newcommand{\eeq}{\end{equation}}

\begin{document}
\title{Electron vortex beams in a magnetic field and spin filter}
\author{ Debashree Chowdhury\footnote{Electronic
address:{debashreephys@gmail.com}}${}^{}$ , Banasri Basu\footnote{Electronic address:{sribbasu@gmail.com}}${}^{}$,  Pratul Bandyopadhyay \footnote{Electronic address:{b$_{-}$pratul@yahoo.co.in}} }
\affiliation{Physics and
Applied Mathematics Unit, Indian Statistical Institute,\\
 203
Barrackpore Trunk Road, Kolkata 700 108, India}


\begin{abstract}
\n
We investigate the propagation of electron vortex beams in a magnetic field. It is pointed out that when electron vortex beams carrying orbital angular momentum propagate in a magnetic field, the Berry curvature associated with the scalar electron moving in a cyclic path around the vortex line is modified from that in free space. This alters the spin-orbit interaction, which affects the propagation of nonparaxial beams. The electron vortex beams with tilted vortex lead to spin Hall effect in free space. In presence of a magnetic field in time space we have spin filtering such that either positive or negative spin states emerge in spin Hall currents with clustering of spin $\frac{1}{2}$ states.

\end{abstract}

\maketitle
~~~~~~~~~~~~~~~$keywords$:~~~   tilted electron vortex beam, Berry curvature, spin filter
\section{Introduction} \label{sec1}
The experimental discovery of electron vortex beams carrying orbital angular momentum \cite{1,2,3} has prompted a good deal of theoretical investigations.  The existence of vortex beams in free space for nonrelativistic scalar electrons was predicted earlier by Bliokh et. al. \cite{4}. The relativistic electron vortex beams representing the angular momentum eigenstates of a free Dirac electron correspond to Bessel beam solutions \cite{5}. Bessel beams in general represent a superposition of monoenergetic plane wave having constant momentum generating a fixed polar angle $\theta_{0}$ with the z axis. In the limit of vanishing spin orbit interaction (SOI) the solutions are eigenstates of both orbital angular momentum (OAM) and spin angular momentum (SAM). These occur for paraxial Besel beams when the polar angle $\theta_{0} \longrightarrow 0.$ However, when the SOI is switched on we have non-paraxial beams which are eigenvalues of the total angular momentum but not of OAM and SAM separately.

It has been shown \cite{7,8} earlier that a fermion can be quantized in the framework of Nelson's stochastic quantization procedure \cite{9,10} when an internal variable is introduced to represent a direction vector (vortex line) attached to a space-time point. The direction vector (vortex line), which is topologically equivalent to a magnetic flux line gives rise to spin degree of freedom and essentially represents a spin vortex.  This effectively gives rise to a gauge theoretical extension of the space-time coordinate as well as momentum and the spin appears as an SU(2) gauge bundle. This framework demonstrates the skyrmionic \cite{11,12} ~ representation of a fermion where it is depicted as a scalar particle encircling the vortex line which is topologically equivalent to a magnetic flux line. 
Interestingly, electron vortex beams carrying OAM appear as a natural consequence  in this skyrmionic representation where scalar electrons move around the vortex line \cite{6}. The geometrodynamics of electron vortex beams is  then governed by the Berry phase \cite{18}acquired by the scalar electron moving around the vortex line. This phase term vanishes when the polar angle $\theta$ formed between the vortex line and the wave front propagation direction ($z$ direction ) is zero 
In fact, this corresponds to the situation when the plane wave vectors of Bessel beams making an angle $\theta_{0}$ with the z axis in the limit of $\theta_{0}\longrightarrow 0.$ In this case, we have paraxial beams which correspond to the electron counterpart of the screw dislocation of optical vortices. When SOI is switched on, caused by the non-zero value of the Berry phase term, nonparaxial beams are generated. Moreover, when the polar angle $\theta$ formed by the vortex line with the z axis is  $\frac{\pi}{2},$ the Berry phase acquired by the scalar electron orbiting around the vortex line involves quantized monopole charge $\mu = \frac{1}{2}$ \cite{6}. This corresponds to the electron counterpart of the edge dislocation of optical vortices. However, for any arbitrary angle ($\theta \neq$ 0, $\frac{\pi}{2}$) the corresponding Berry phase involves non-quantized monopole charge and corresponds to the mixed edge-screw dislocation of optical vortices. Bessel beams in this case involve tilted vortices with respect to the wavefront propagation direction. It has been argued that for the propagation of electron vortex beams in free space with tilted vortices, we have spin Hall effect \cite{6}.

Our goal here is to analyze the situation when the electron vortex beams move in an external magnetic field. Indeed, this external field will have effect on  the Berry curvature associated with the cyclic motion of the scalar electron moving around the vortex line. The change in the Berry phase modifies the SOI and as such have impact on the  nonparaxial beams. In a magnetic field the Berry phase is modified by the Gouy phase factor which is associated with the diffractive Laguerre-Gaussian (LG) beams in free space.
In a magnetic field this is related to the transverse kinetic energy of spatially confined modes and leads to the contribution of the Gouy energy to Landau energy \cite{14}.
 This phase factor determines the squared $spot~ size$ of the LG beams \cite{15}. It is observed that this change in the effective Berry phase in an external magnetic field may alter significantly the situation related to the generation of the spin Hall currents in electron vortex beams with tilted vortices. Indeed, in the presence of an external magnetic field in time space we have spin filtering and clustering of spins. 
 However, if the external field is considered to be a hopping one, such that at two consecutive time sequences there is a flip in the orientation of the field in two opposite directions we have alternating spin Hall currents with positive and negative spins.

In sec. II we shall recapitulate certain features of skyrmionic model of a fermion and its relevance in electron vortex beams. Sec. III deals with the study of  the spin-orbit couplings in an external field. In sec.IV we narrate  the situation related to spin Hall effect when vortex beams with tilted vortices move in a magnetic field in time space.

\section{Skyrmionic model of a fermion and electron vortex beams}
The quantization of a fermion can be achieved \cite{7,8} in the framework of Nelson's stochastic quantization procedure \cite{9,10} when an internal variable is introduced to represent a direction vector, giving rise to the spin degrees of freedom. This effectively gives rise to the SL(2,C) gauge theory and demanding Hermiticity the gauge field belongs to the SU(2) group. The spin degrees of freedom appear as an SU(2) gauge bundle. This represents a gauge theoretical extension of the space-time coordinate as well as momentum, which can be written as gauge covariant operator acting on functions in phase space
\begin{eqnarray}\label{eq1}
Q_\mu &= & -i\left( \frac{\partial}{\partial p_\mu}+{\cal{A}}_\mu (p)\right) \nonumber \\
P_\mu &= & i\left( \frac{\partial}{\partial q_\mu}+{\cal{B}}_\mu (q)\right)
\end{eqnarray}
where ${\cal{A}}_\mu ({\cal{B}}_\mu)$ is the momentum (spatial coordinate) dependent SU(2) gauge field. Here $q_\mu(p_\mu)$ denotes the mean position (momentum) of the external observable space. In this scenario a massive fermion appears as a skyrmion.
 In fact as the direction vector represents a vortex line which is topologically equivalent to a magnetic flux line, we may depict a fermion as a scalar particle orbiting around a magnetic flux line. When a particle encircles the loop enclosing the magnetic flux line it acquires a geometric phase (Berry phase) apart from the usual dynamical one \cite{18}. The Berry phase acquired by the scalar particle is given by 2$\pi \mu$ where $\mu$ is the monopole charge associated with the magnetic flux line \cite{19}. When the monopole is located at the center of a unit sphere the Berry phase is given by $\mu\Omega(C)$ where $\Omega(C)$ is the solid angle subtended by the closed contour at the origin and is given by \beq \Omega(C) = \int_{C}(1 - cos \theta)d\phi = 2\pi(1- cos\theta),\label{eq8}\eeq where $\theta$ is the polar angle of the vortex line with the quantization axis(z axis). So for $\mu = \frac{1}{2},$ corresponding to one magnetic flux line the Berry phase is
\beq \phi_{B} = \pi(1 - cos\theta)\label{eq9}.\eeq Transforming to a reference frame where the scalar electron is considered to be fixed and the vortex state (spin state) moves in the field of the magnetic monopole around a closed path, $\phi_{B}$ in equation (\ref{eq9}) corresponds to the geometric phase acquired by the vortex state. The angle $\theta$ corresponds to the deviation of the vortex line from the $z$ axis. Equating $\phi_{B}$ in (\ref{eq9}) with $2\pi\mu$ which is the geometric phase acquired by the scalar electron moving around the vortex line in a closed path, we find the associated effective monopole charge as
\beq \mu = \frac{1}{2}(1 - cos\theta)\label{eq12}.\eeq
This suggests that for $\theta = 0$ and $\frac{\pi}{2}$, $\mu$ takes quantized values but for other angles $0 < ~\theta<~\frac{\pi}{2}$ it is non-quantized.
When the vortex line representing the spin axis is  parallel to the wave propagation direction implying $\theta =$ 0 we have the paraxial vortex beams whereas when $\theta = ~ \frac{\pi}{2}$ the vortex line is orthogonal to the wave propagation direction. For other values of $\theta,$ corresponding to non-quantized monopole charge, the vortex line is tilted in an arbitrary direction. Actually, these angular values of $\theta$ correspond respectively to the screw, edge and mixed screw-edge dislocations of optical vortices.

\section{electron vortex beams in a magnetic field and spin-orbit interaction}
In this section we consider the electron vortex beam in presence of an external magnetic field ($\vec{B}$). If the magnetic field be axially symmetric longitudinal one, the vortex vector potential can be chosen as \cite{14} \beq \vec{A}({\vec r}) = \frac{B(r)r}{2}\hat{\vec{e}}_{\phi},\eeq where $\vec{B} = \vec{\nabla}\times\vec{A}.$
Now if we consider the electron vortex beams in  presence of a magnetic field oriented along the z axis having magnetic flux $\phi$, we can write the vortex vector potential as  
\beq \vec{A}({\vec r}) = 
 \frac{\phi}{2\pi r}\hat{\vec{e}}_{\phi} = \frac{\alpha}{r}\hat{\vec{e}}_{\phi},\eeq with  $\alpha = \frac{\phi}{2\pi}$ being denoted as the magnetic field parameter. The non-relativistic Hamiltonian with covariant momentum is given by
\beq H = \frac{1}{2m}(\vec{p} - e\vec{A}({\vec{r}}))^{2}.\eeq 
In the non-relativistic case, the configuration variables can be obtained from eqn. (1) in the sharp point limit \cite{pb}. From eqn. (1), the gauge theoretical extended spatial coordinate can be written as $\vec{R} = \vec{r} + \vec{{\cal A}}(p),$ where $\vec{{\cal A}}(p)$ is the SU(2) gauge field representing  the spin degrees of freedom. In the sharp point limit the Schroedinger equation in cylindrical coordinates, in terms of the external observable variable, is given by \cite{14}
\beq -\frac{\hbar^{2}}{2m}\left[\frac{1}{r}\frac{\partial}{\partial r}\left(r\frac{\partial}{\partial r}\right) + \frac{1}{r^{2}}\left(\frac{\partial}{\partial \phi} + ig\frac{2r^{2}}{w_{m}^{2}}\right)^{2} + \frac{\partial^{2}}{\partial z^{2}}\right]\psi = E\psi,\label{4}\eeq
where $w_{m} = 2\sqrt{\frac{\hbar}{|eB|}}$ is the magnetic length parameter, $g = sgn B = \pm 1$ indicates the direction of the magnetic field. The parameter $\frac{2r^{2}}{w_{m}^{2}}$ essentially represents the magnetic field parameter $\alpha.$ 

To study the effect of magnetic field on different spin related issues of electron vortex beams through Berry phase we would like to find out the SOI first as the Berry phase can be effectively studied \cite{6, 13} via SOI. This SOI is evaluated from the modified OAM and is given by 
\beq \vec{{\tilde L}} = \vec{R}\times \vec{P},\eeq where $\vec{R} = \vec{r} - \vec{{\cal A}}(p)$ is the modified coordinate with  $\vec{{\cal A}}(p)$ is the SU(2) gauge field and $\vec{P} = \vec{p} - \vec{A}(r),$ is the covariant momentum which incorporates the gauge potential $\vec{A}(r)$ corresponding to the external field.

 One may note that the SOI in the non-relativistic limit can be derived from the Dirac equation by introducing Foldy-Wouthuysen transformations separating the positive and negative energy components of the Dirac equation. Using the projection on the positive energy subspace, the electron position operator is given by \cite{2o, 2t}
\beq \vec{R} = \vec{r} - \vec{{\cal A}}^{~'}(p),\eeq with
\beq {\cal{\vec{A}}}^{~'}(\vec{p}) = \frac{\vec{p}\times\vec{\sigma}}{2p^{2}}\left(1 - \frac{m}{E}\right)\label{143}.\eeq 
In the relativistic case $\frac{m}{E} \rightarrow 0$ and in this case we realize the spatial coordinate as $\vec{R} = \vec{r} - \vec{{\cal A}}(p)$ from eqn (1) with 
\beq {\cal{\vec{A}}}(\vec{p}) = \mu\frac{\vec{p}\times\vec{\sigma}}{p^{2}},\label{1w}\eeq where $\mu$ represents the monopole strength having the value $|\mu| = \frac{1}{2}$ and $\vec{\sigma}$ is the vector of Pauli matrices. We may mention here that from the connection (11), we obtain the Berry phase by integrating it along the contour of the Bessel beam spectrum in momentum space given by \cite{5}
\beq \phi_{B} = \int  {\cal{\vec{A}}}^{~'}(\vec{p}) d\vec{p} = 2\pi \Delta s,\eeq where $\Delta s$ is the spin variable which induces the SOI by modifying the OAM. 
As discussed in the previous section, in the present formalism the Berry connection corresponds to a monopole and the Berry phase acquired by the scalar particle orbiting around the vortex line (magnetic flux line) in a closed loop is given by $2\pi \mu$ \cite{19}.

 The angular momentum of a charged particle in the field of a magnetic monopole is given by \beq \vec{J} = \vec{L} - \mu\hat{\vec{r}},\eeq where $\vec{L}$ is the OAM. Thus when OAM is vanishing, the total angular momentum $\vec{J}$ which is effectively the spin is given by $|\mu|$ with $S_{z} = \pm \mu.$ In view of this, we note that the Berry phase obtained in terms of the spin variable $\Delta s$ in Dirac equation formalism corresponds to the phase obtained in terms of the monopole charge $\mu.$ 
 
 The SOI in presence of a magnetic field can be obtained from the modified OAM operator and can be written as 
\begin{eqnarray}
 \vec{\tilde{L}} &=& \vec{R} \times \vec{P}\nonumber\\
 &=& \left(\vec{r} - \vec{{\cal {A}}}(p)\right) \times \left(\vec{p} - \vec{A}(r)\right)\nonumber\\
 &=& \vec{r}\times \vec{p} - \vec{{\cal {A}}}(p)\times \vec{p} - \vec{r}\times \vec{A}(r) +  \vec{{\cal {A}}}(p)\times \vec{A}(r)\nonumber\\
 &=& \vec{L} - \vec{L}_{1} - \vec{L}_{2} + \vec{L}_{3}.
\end{eqnarray}
To find $\langle\vec{\tilde{L}}\rangle$ we have to  calculate $\langle \vec{L}_{1}\rangle, \langle \vec{L}_{2}\rangle$ and $\langle \vec{L}_{3}\rangle$ explicitly.

From eqn. (12) we can write
\beq \langle \vec{L}_{1}\rangle = \vec{{\cal {A}}}(p)\times \vec{p} = \langle \mu \frac{\vec{p}\times \vec{\sigma}\times\vec{p}}{p^{2}} \rangle = -\langle \mu\vec{p}\times(\frac{\vec{p}}{p^{2}}\times\vec{\sigma}) \rangle.\eeq The expectation value of $\vec{\sigma}$ is given by
\beq \langle\vec{\sigma}\rangle = \frac{\langle\psi|\vec{\sigma}|\psi\rangle}{\langle\psi|\psi\rangle} = \vec{n}\label{26},\eeq
where $\psi$ is a two-component spinor
$ \psi = \left(\begin{array}{cr}
   \psi_{1} \\
    \psi_{2}
 \end{array}\right) \label{27}$ with $\langle\psi|\psi\rangle = 1$ and 
  $\vec{n}$ is a unit vector. Thus, 
  \beq \langle\vec{L}_{1}\rangle = -\langle \mu\vec{p}\times(\frac{\vec{p}}{p^{2}}\times\vec{\sigma})\rangle = -\langle\mu \vec{\kappa}\times (\vec{\kappa}\times \vec{\sigma})\rangle\label{25}\eeq with $\frac{\vec{p}}{p} = \vec{\kappa},$ $\vec{\kappa}$ being the unit vector. This gives  \beq \langle\vec{L}_{1}\rangle = -\mu \vec{n}\label{29}.\eeq Similarly, for $\vec{L}_{2}$ we have
\begin{eqnarray}
\vec{L}_{2} &=& \vec{r}\times \vec{A}(r)\nonumber\\
&=&  \vec{r}\times \frac{\alpha}{r}\hat{\vec{e}}_{\phi} \nonumber\\
&=& -\alpha\vec{\tilde{\kappa}}\times \hat{\vec{e}}_{\phi},
\end{eqnarray}
where $\vec{\tilde{\kappa}} = \frac{\vec{r}}{r}.$ This gives \beq \langle\vec{L}_{2}\rangle = -\langle\alpha\rangle\vec{\tilde{n}}, \eeq
$\vec{\tilde{n}}$ being another unit vector.
Because of the fact that spin and orbital angular momentum are orthogonal to each other, we have $\vec{\sigma}\times\vec{r} \times\vec{p} = ~0.$ With this, we can write
\begin{eqnarray}
\vec{L}_{3} &=&  \vec{{\cal {A}}}(p)\times \vec{A}(r)\nonumber\\
&=& 0.
\end{eqnarray}
The term $\vec{L}_{3}$ contributes nothing to the OAM and thus we have $\langle\vec{L}_{3}\rangle = 0.$ Finally, \beq \langle\vec{\tilde{L}}\rangle = \langle\vec{L}\rangle -\langle\vec{L}_{1}\rangle - \langle\vec{L}_{2}\rangle = l+\mu+ \langle\alpha\rangle,\eeq where
$\langle \alpha\rangle$ is given by \beq \langle \alpha\rangle = \left\langle \frac{2r^{2}}{w_{m}^{2}}\right\rangle = \frac{\left\langle\psi |\frac{2r^{2}}{w_{m}^{2}}|\psi\right\rangle}{\left\langle\psi |\psi\right\rangle}.\eeq
From the solution of eqn (8), which has the form of nondiffracting Laguerre-Gaussian (LG) beams given by
\beq \psi^{L}_{l,n} \simeq \left(\frac{r}{w_{m}}\right)^{|l|}L_{n}^{|l|}\left(\frac{2r^{2}}{w_{m}^{2}}\right)exp\left(-\frac{r^{2}}{w_{m}^{2}}\right)exp\left[i(l\phi+k_{z}z)\right],\label{6}\eeq
 we readily obtain  \cite{14}
\beq \langle \alpha\rangle = \langle \frac{2r^{2}}{w_{m}^{2}}\rangle = 2n + |l| + 1\eeq
where $n=0,1,2....$ is the radial quantum number and $|l|$ is the azimuthal quantum number. 
The modified OAM is thus obtained as
\begin{eqnarray}
\langle\vec{\tilde{L}}\rangle 
&=& l + \mu + g~ \langle \alpha\rangle \nonumber\\
&=& l + \mu + g~ (2n + |l| + 1),
\end{eqnarray}
where $g$ denotes the sign of the external magnetic field.
This suggests that a part of the angular momentum is transformed from the SAM to OAM implying SOI.

With all these expressions at hand eqn. (8) helps us to write the dispersion relation ($\hbar = 1$)   as 
\beq E = \frac{p^{2}}{2m} - \Omega l + |\Omega|(2n+|l|+1) = E_{\lVert} + E_{Z} + E_{G}.\eeq Here $E_{\lVert} = \frac{p^{2}}{2m}$ is the energy of the free longitudinal motion and the quantized transverse motion energy corresponds to the term \beq E_{\perp} = E_{Z} + E_{G},\eeq with $E_{Z}$ and $E_{G}$ representing Zeeman and Gouy energy respectively. 
Eqn.(28)  displays that the Berry phase of the scalar electron orbiting around the vortex line in presence of a magnetic field is modified by the Gouy phase factor $\langle \alpha\rangle,$ which is related to the diffractive LG beam in free space. In a magnetic field the Gouy phase is related to the contribution due to the transverse motion energy $E_{G}$ in eqn (28). This phase factor essentially determines the squared spot size of the LG beam \cite{14}.

The modified Berry phase is now given by \beq \phi_{B} = 2\pi(\mu + g\langle\alpha\rangle ) = 2\pi(\mu + g(2n + |l| + 1)),\eeq where $g = \pm 1,$ represents the sign factor depending on the orientation of the magnetic field. It is noted that $\langle\alpha\rangle = (2n + |l| + 1)$ being an integer will contribute trivially to the phase factor and so the effective phase is given by $2\pi\mu.$ 
Though the external magnetic field changes the Berry phase trivially, the change in the Berry curvature will have its effect on the anomalous velocity which is associated with the spin Hall effect generated in the vortex beams with tilted vortices, which  will be  discussed in the next section. 

\section{Tilted vortex and spin Hall effect in an external magnetic field in time space}
In a recent paper \cite{6} it has been pointed out that electron vortex beams in free space with tilted vortex give rise to spin Hall effect. This follows from the fact that, as indicated in sec II, for tilted vortex the associated Berry phase acquired by the scalar electron moving around the vortex line involves nonquantized monopole charge which we denote as $\tilde{\mu}$. The non-quantized monopole charge $\tilde{\mu}$ undergoes a renormalization group flow and can be considered as a time dependent parameter \cite{20,21}.
The time dependence of the monopole charge makes the corresponding gauge field a time dependent one. The time derivative of the gauge field  $\frac{\partial\vec{A}}{\partial t}$ generates an electric field $\vec{{\cal E}}$ which accelerates electrons so that the momentum carries explicit time dependence. We denote the time dependent momentum as $\vec{k}.$ In this case we can introduce a non-inertial coordinate frame with basis vectors $(\vec{f}, \vec{w}, \vec{u})$ attached to the local direction of momentum $\vec{u} = \frac{\vec{k}}{k}.$ This coordinate frame rotates as $\vec{k}$ varies with time. Such rotation with respect to a motionless (laboratory) coordinate frame describes a precession of the triad $(\vec{f}, \vec{w}, \vec{u}),$ with some angular velocity. When we take the direction of the vortex line at an instant of time as the local $z$ axis which represents the direction of propagation of the wave front, we note that this corresponds to the paraxial beam in the local frame. In this local non-inertial frame, the local monopole charge will correspond to a pseudospin. Indeed, the expectation value of the spin operator
\beq \left\langle \vec{S}\right\rangle = \frac{1}{2}\frac{\langle\psi|\vec{\sigma}|\psi\rangle}{\langle\psi|\psi\rangle}\label{43},\eeq
undergoes precession with the precession of the coordinate frame. This implies that the polarization state depends on the choice of the coordinate frame \cite{23}. When the direction of the vortex line is taken to be the local $z$ axis in the non-inertial frame, the local value of $\tilde{\mu}$ is changed and takes the quantized value $|\mu| = \frac{1}{2}$ owing to the precession of the spin vector and thus corresponds to the pseudospin  in this frame. The pseudospin vector is parallel to the momentum vector $\vec{k}.$ In terms of the time dependent momentum $\vec{k}$, the Berry curvature is given by
\beq \vec{\Omega}(\vec{k}) = \mu\frac{\vec{k}}{k^{3}}.\label{40}\eeq  This curvature gives rise to an anomalous velocity
\beq \vec{v}_{a} = \dot{\vec{k}}\times \vec{\Omega}(\vec{k}) = \mu \dot{\vec{k}}\times\frac{\vec{k}}{k^3}. \label{41}\eeq
 Thus the anomalous velocity is perpendicular to the pseudospin vector and points along the opposite directions depending on the chirality  $s_{z} = +\frac{1}{2}(-\frac{1}{2})$ corresponding to $\mu> 0(< 0).$ This separation of the spins gives rise to the spin Hall effect. It is observed that $\mu$ here corresponds to the helicity as we have $\mu = s_{z} = \pm \frac{1}{2}.$ Denoting $\frac{\dot{\vec{u}}}{|\dot{\vec{u}}|} = \vec{n},$ we note that the spin current is orthogonal to the local plane ($\vec{u}, \vec{n}$). Thus we find that for a tilted vortex with respect to the wave propagation direction, though the Berry phase acquired by the orbiting scalar electron may be viewed as an artefact of a rotating coordinate frame, the spin Hall effect is a Coriolis type transverse deflection as the spin current is orthogonal to the local plane $(\vec{u},\vec{n})$.

 When we consider electron vortex beams with tilted vortex in an external magnetic field in time space we note that the corresponding time dependent gauge potential will modify the Berry curvature given by (32) . 
The modified  Berry curvature 
is given by $\tilde{\vec{\Omega}} = \vec{\Omega}(\vec{k}) + \vec{B}(\vec{k}),$ where $\int_{\Sigma}\vec{B}(\vec{k})d\vec{\Sigma}$ is the contribution to the Berry phase due to the external field and represents the flux passing through the surface $\Sigma$ in momentum space. 

The Berry curvature of  the electron vortex beams propagating in a magnetic field is then given by  
 \beq \tilde{\vec{\Omega}}(\vec{k}) = (\mu + g\langle\alpha\rangle) \frac{\vec{k}}{k^{3}},\label{44}\eeq
 where $g\langle\alpha\rangle \frac{\vec{k}}{k^{3}}$ is the contribution of the external field to the Berry curvature in momentum space and  $g = \pm 1$ denotes the sign factor depending on the orientation of the magnetic field.  
 The  anomalous velocity is then given by
 \beq \vec{\tilde v}_{a} = \left(\mu + g\langle\alpha\rangle\right) \frac{\dot{\vec{k}}\times\vec{k}}{k^{3}} =\left[\mu + g(2n + |l| + 1)\right] \frac{\dot{\vec{k}}\times\vec{k}}{k^{3}}.\eeq 
	with $\mu = \pm\frac{1}{2}$ in the local non-inertial frame and $n$ and $|l|$ are integers. 
	It may be mentioned that, Fujita at el \cite{25} have shown that a well defined gauge field in time space which has the physical significance of an effective magnetic field is found to be the underlying origin of the anomalous velocity owing to the curvature in momentum space. In our present study we find that the Berry curvature associated with the orbiting scalar electron is modified by the curvature generated by the time dependent magnetic field and hence alters the anomalous velocity in free space.
  As mentioned in Sec. III the quantized value of $|\mu|= \frac{1}{2}$ corresponds to the spin with $s_z=\pm\frac{1}{2},$ the modified value of the Berry phase factor $|\mu+g\langle\alpha\rangle|$ corresponds to the total spin of the system. The resultant spin of the system is now given by $|\mu + g\langle\alpha\rangle| =| \mu + g (2n + |l|+1)|.$  
	The factor $\langle\alpha\rangle = (2n + |l| + 1)$ together with $\mu =\pm \frac{1}{2}$ always give rise to the clustering of spin $\frac{1}{2}$ states. For the lowest energy state with $n=0$ and for positive $g$, value of g$\langle\alpha\rangle = ~1.$  
 For $\mu = \frac{1}{2}$ (-$\frac{1}{2}$), we have the helicity states as $\mu + \langle\alpha\rangle =~ \frac{3}{2}$ ($\frac{1}{2}.$) For higher values of $n$ and $|l|$, we will have higher half integer values of ($\mu + \langle\alpha\rangle$). 
One point may be noted that for $+\langle\alpha\rangle$ we will always have up spins. Similarly, for $-\langle\alpha\rangle$ we will have only down spins. Thus depending on the orientation of the magnetic field there is either positive or negative spin current moving in opposite direction which leads to spin filtering. Our analysis suggests that a time variant magnetic field can be used to have spin filtering of the tilted electron vortex beams.

In a different approach, Karimi et. al. in \cite{Karimi} have proposed a space variant Wein filter where the geometric phase plays a crucial role. They have argued that when electron vortex beams are subjected to a space variant magnetic field and a suitable electric field one can conceive an apparatus as a space variant Wein filter which induces a spin half turn. This can be used for generating a pure electron vortex beam from a spin polarized beam and conversely a spin polarized beam from a pure electron vortex beam. This is caused by the conversion of SAM and OAM through the generation of the geometric phase which arises because of the spin manipulation.

Eqn. (35) suggests that when an external time dependent magnetic field is introduced such that electron vortex beams with tilted vortex propagate in this field we can tune the spin Hall current. As the expression of spin current is given by
\beq j^{k}_{s,j} = \frac{\hbar}{4}\left\langle \{\sigma^{k}, {\tilde v}_{a,j}\}\right\rangle, 
\eeq we can have in the system spin currents with either positive or negative spin states with clustering of odd number of spin $\frac{1}{2}$ states. 
Importantly, if the external field is of hopping type so that at two consecutive time sequences the orientation of the field is altered, we have alternating spin Hall current such that the spin currents with $s_{z} = \pm\frac{n}{2},$ where n = 1,3,5....., propagate in opposite directions  in consecutive time sequences.
In a nutshell, propagation of tilted electron vortex beams in an external time dependent magnetic field gives rise to polarization of  spins such that we have spin currents as well as clustering of spin 1/2 states. This theoretical analysis can be used to demonstrate a  spin filter such that at the output we have spin currents with either for up spin or for down spin of the cluster. 

\section{Discussion}
We have studied here the situation when electron vortex beams propagate in an external magnetic field from the prospective of the skyrmion model of a fermion. In fact the skyrmion model is achieved through the quantization of a Fermi field in the framework of Nelson's stochastic quantization procedure when we introduce a direction vector (vortex line) attached to the space-time point, which gives rise to the spin degrees of freedom. A vortex line is topologically equivalent to a magnetic flux line. Thus the skyrmionic picture of a fermion depicts it as a scalar particle moving around the magnetic flux line.  The electron vortex beam is a natural consequence of this picture of a fermion. In a recent paper \cite{6} it has been observed that the dynamics of electron vortex beams is determind by the Berry phase acquired by the scalar electron orbiting around the vortex line. Indeed it is found that when the phase involves quantized Dirac monopole we have paraxial (nonparaxial) beams such that the vortex line is parallel (orthogonal) to the wave front propagation direction. However when the Berry phase involves nonquantized monopole charge we have tilted vortex with respect to the wave front propagation direction. It has been pointed out that nonparaxial beams involve SOI which appears as a manifestation of the Berry phase. Electron vortex beams with tilted vortex gives rise to spin Hall effect in free space.

In the present study we show that in presence of an external magnetic field in time space the Berry curvature is changed and the phase is modified by the magnetic field parameter. This parameter is associated with the Gouy phase which is related to the diffractive LG beams in free space. In a magnetic field, the Gouy phase factor is associated with the transverse kinetic energy of spatially confined mode and  and leads to the contribution of the Gouy energy to Landau energy. Indeed, this parameter determines the squared $spot size$ of the LG beam. It has been argued here that when electron vortex beams propagate in an external time dependent magnetic field we have modification of the spin Hall current from that in the free space for tilted vortex. It is found that this leads to spin filtering when either positive or negative spin Hall currents appear depending on the orientation of the magnetic field with clustering of spins. For a hopping field which alters its orientation at two consecutive time sequences, we have alternating spin Hall currents with positive and negative spin states at successive time sequences.

\begin{center}
{\bf Acknowledgment}
\end{center}

We would like to acknowledge the anonymous referee for his/her valuable comments. 

\end{document}